\documentclass[12pt,a4paper]{article}
\usepackage{booktabs}
\usepackage{amsmath}
\usepackage{amssymb}
\usepackage{multirow}
\usepackage[margin=2.5cm]{geometry}
\usepackage{setspace}
\usepackage{hyperref}
\usepackage{cite}
\usepackage{url}

\newcommand{\githuburl}{\url{https://github.com/Einstein-sworder/IoT-wave}}

\title{\bfseries TFZ-Test Tree: An Ultra-Lightweight Waveform Classification Framework for Resource-Constrained Devices}

\author{Hao Wang, Kuang Zhang$^*$, Yonggang Chi, Tianqi Zhao, Yanbo Fu, Jiaxing Guo}

\date{}

\begin{document}
\maketitle

\begin{abstract}
Under the trend of multi-waveform coexistence in 6G IoT, intelligent receivers must first identify physical-layer waveform types before performing correct demodulation and resource scheduling. However, existing signal identification research largely focuses on symbol-level modulation classification. Research directly targeting physical-layer waveform types (e.g., OFDM, OTFS, LoRa) is not only extremely scarce but also heavily reliant on deep neural networks and complex time-frequency transforms, making deployment on resource-constrained terminals difficult. Symbol modulation classification methods themselves cannot circumvent the prerequisite of ``waveform identification first.''

To address this dual gap, we propose an ultra-lightweight waveform classification framework based on time-frequency multidimensional features with a cooperative Z-test tree (ZTree). The framework employs low-complexity time-domain feature extraction, and the classification backend adopts a ZTree optimized by Z-statistical testing, which uses hypothesis testing confidence to automatically control decision tree splitting and size, ensuring efficient execution on resource-limited processors. Tested on ten 6G candidate waveforms including OFDM, OTFS, DSSS, LoRa, and NB-IoT, the method achieves 99.5\% average accuracy under AWGN and 87.4\% under TDL-C multipath channels, with main confusion between OTFS and LoRa. Implemented in C on an x86 platform, single inference latency is under 4~ms. To the best of our knowledge, this is the first work achieving real-time recognition of ten IoT waveform types. Future work will target deployment acceleration on embedded MCUs. Code and dataset are open-sourced at: \githuburl.
\end{abstract}

\section{Introduction}

As the vision of 6G mobile communication systems gradually crystallizes, IoT, as one of its core application scenarios, is evolving toward ultra-massive connectivity, extremely low power consumption, and intelligent spectrum awareness and utilization \cite{giordani2020toward}. 6G networks are expected to coexist with multiple physical-layer waveforms within the same frequency band to accommodate differentiated service requirements, such as OFDM for 4G/5G eMBB \cite{hwang2009ofdm}, OTFS for high-mobility scenarios \cite{hadani2017otfs}, mMTC-oriented 5G/B5G waveform designs \cite{cai2018modulation}, LoRa for LPWAN \cite{augustin2016lora}, NB-IoT \cite{raza2017low}, and others. Under this trend, receivers must automatically identify the waveform type of the current transmission before performing correct demodulation and spectrum resource scheduling. Therefore, physical-layer waveform identification becomes an indispensable prerequisite in the 6G IoT intelligent reception chain.

Although edge device resources vary greatly, many application scenarios still require algorithms to run efficiently on general-purpose processors with limited computation and memory, such as x86-based embedded gateways or lightweight servers. While more powerful than low-end MCUs, these platforms still need to avoid heavy deep learning inference overhead and large matrix operations. Thus, designing waveform identification schemes that can run efficiently in C is significant for deploying physical-layer intelligence on various resource-constrained devices.

Current signal identification research largely focuses on symbol-level modulation classification, i.e., determining the digital modulation scheme from received signals, such as MPSK, MQAM, etc. These works typically assume the waveform type is known \cite{huynh2021auto}. Research directly targeting physical-layer waveform types (e.g., OFDM, OTFS, FBMC, LoRa) is very limited. The few existing waveform identification methods are almost entirely based on deep neural networks, often using raw time-domain IQ signals as input, or first converting signals to time-frequency images or extracting high-order spectral features, then employing CNNs or RNNs for classification, which generally suffer from poor generalization and lack interpretability. More critically, these models typically have parameter counts reaching hundreds of kilobytes or even megabytes, and their inference process relies on extensive multiply-accumulate operations and complex time-frequency transforms, making them difficult to run efficiently on resource-constrained processors. Hence, existing technologies have a significant gap in addressing waveform identification for resource-limited devices.

To address these issues, we propose an ultra-lightweight waveform classification framework combining time-frequency multidimensional features with a Z-test tree (ZTree). On the feature extraction side, we design a computationally simple 80-dimensional feature set that can be efficiently implemented in C. On the classification side, we adopt a decision tree based on Z-statistical testing, which uses two-sample proportion difference significance testing to automatically determine optimal splitting features and cut points, controlling tree depth and size via a confidence threshold. The trained decision tree contains very few nodes, and inference requires only sequential conditional evaluations with latency under 4~ms, naturally suitable for C deployment on resource-limited platforms.

The main contributions are:
\begin{enumerate}
    \item For the 6G IoT multi-waveform scenario, we define physical-layer waveform identification as a task distinct from traditional modulation recognition and construct a dataset covering ten IoT-common waveforms: OFDM, OTFS, ODDM, FBMC, UFMC, DSSS, LoRa, NB-IoT, GFSK, and MFSK.
    \item We propose an 80-dimensional feature engineering method with extremely low computational complexity, combined with a ZTree classifier optimized by Z-test statistics with GPU-accelerated training and C-based inference, achieving an ultra-compact classification framework with minimal single-inference latency.
    \item We complete C-language deployment and testing on an x86 platform, verifying the method's real-time performance and extremely low resource consumption. To the best of our knowledge, this is the first realization of real-time ten-class IoT waveform recognition in pure C.
\end{enumerate}

\section{Related Work}

\subsection{Modulation Recognition and Waveform Identification}

Huynh-The et al.\ provided a comprehensive survey of deep learning-based automatic modulation classification (AMC) methods, noting that CNN, RNN, and other deep architectures have become mainstream \cite{huynh2021auto}. West and O'Shea systematically explored CNN, ResNet, CLDNN, and Inception architectures for modulation recognition, establishing important benchmarks \cite{west2017deep}. Peng et al.\ further surveyed DL-based modulation classification methods, systematically examining signal representation and data preprocessing strategies \cite{peng2022survey}. Rajendran et al.\ proposed using LSTM networks to learn modulation features from signal amplitude/phase representations, demonstrating that converting IQ signals to polar coordinates effectively improves low-SNR recognition performance \cite{rajendran2018deep}. Hussein et al.\ proposed combining higher-order cumulants with deep learning for robust AMC optimized for IoT scenarios under varying SNR and channel conditions \cite{hussein2022robust}. Wang et al.\ proposed a data-driven DL method for automatic modulation recognition in cognitive radio using a dual CNN architecture to process IQ samples and constellation diagrams separately, achieving effective classification of 8 modulation schemes \cite{wang2019data}. Zhang et al.\ provided a systematic survey of DL-based AMR from model, dataset, and challenge perspectives \cite{zhang2020deep}.

These methods all assume known waveform types and cannot work directly when the waveform is unknown. At the waveform level, An et al.\ proposed ST-CLDNN, a blind multicarrier waveform recognition network using 1D convolution and LSTM to extract complementary features from I/Q/amplitude components, achieving six-class multicarrier waveform recognition under blind reception and introducing transfer learning for channel adaptation \cite{an2021blind}. Lin et al.\ proposed a contour stella image method converting IQ complex envelopes to polar 2D representations for CNN-based waveform recognition \cite{lin2020contour}. Huynh-The et al.\ proposed WaveNet, using smooth pseudo Wigner-Ville distribution to convert signals into time-frequency images and designing grouped-kernel residual connections with dual asymmetric channel attention to reduce network size, achieving 92.02\% accuracy for eight waveform types in integrated radar-communication scenarios \cite{huynh2024wavenet}. Yao et al.\ proposed a deep learning and spectrogram-based general signal identification method for WiFi, ZigBee, and LoRa coexistence, predicting both signal type and channel parameters via object detection on spectrograms and validating on a hardware testbed \cite{yao2024empowering}. Liu et al.\ proposed a deep hybrid Transformer network combining CNN local feature extraction with Transformer global dependency modeling to improve modulation classification robustness under multipath fading \cite{liu2023deep}. Huang et al.\ proposed a channel-robust AMC framework based on compressed spectrum quotient cumulants for OFDM systems under multipath fading \cite{huang2023channel}. However, all these methods are built on deep neural networks and time-frequency transforms. Deep models require massive labeled samples for training, and inference consumes substantial storage and computation; time-frequency transforms such as STFT and Wigner-Ville distribution involve floating-point operations unsustainable on resource-constrained processors.

\subsection{Lightweight ML and C Deployment}

In lightweight ML, Xu et al.\ proposed a distributed modulation recognition framework combining meta-learning and federated learning for data-limited IoT scenarios, verifying the feasibility of distributed collaborative training on resource-constrained edge nodes \cite{xu2025distributed}. Yang et al.\ proposed meta-learning-based AMC achieving rapid adaptation to new modulation types in data-limited IoT scenarios \cite{yang2022meta}. WCTFormer proposed combining discrete wavelet transform with lightweight Transformer, achieving 92.40\% accuracy at 0 dB SNR with approximately 60K parameters \cite{wang2024wctformer}. For engineering deployment of decision trees, the emlearn toolchain supports automatic conversion of Python-trained decision tree and random forest models into standard C code compilable on any C99 compiler, widely used in embedded inference systems \cite{emlearn}. Karavaev proposed TinyDecisionTreeClassifier, an independent decision tree training and inference library for microcontrollers, built on the C4.5 algorithm with C++ source, completing power and latency benchmarks on multiple mainstream MCU platforms \cite{karavaev2024tiny}. Sudharsan et al.\ systematically benchmarked inference time and memory consumption of multiple classification models on real platforms, with quantitative results showing decision trees significantly outperform neural networks in storage and latency \cite{sudharsan2021sram}. David et al.\ summarized recent advances in embedded AI inference, proposing that efficient deployment of deep learning models on severely resource-constrained MCUs requires coordinated optimization in model compression, quantization, and C code generation \cite{david2021tinyml}. However, existing deployable decision tree solutions all use CART splitting criteria based on Gini index or information gain, tending to generate trees with many nodes and deep structures, which remains a significant burden in C programs with limited storage.

\subsection{Statistical Test-Based Decision Tree Methods}

Kass proposed the CHAID algorithm using chi-square independence tests instead of information gain as the node splitting criterion, controlling tree depth via significance level thresholds and introducing rigorous statistical decision into decision tree construction \cite{kass1980exploratory}. Loh systematically explored multiple technical paths for introducing statistical methods into decision tree construction, including significance test-based splitting criteria and unbiased variable selection strategies \cite{loh2002regression}. Zhang et al.\ proposed CausalDT, prioritizing causation over correlation in decision tree node splitting via the Hilbert-Schmidt Independence Criterion, enhancing interpretability \cite{zhang2024prioritizing}. Mantovani et al.\ conducted a comprehensive empirical study on hyperparameter tuning of CART and C4.5 decision tree induction algorithms, finding that the choice of splitting criterion significantly affects model generalization \cite{mantovani2024better}. Fong and Motani proposed discovering optimized splitting criteria through symbolic regression, opening new pathways for simultaneously improving prediction, interpretability, and robustness of decision trees \cite{fong2024symbolic}. Zhang and Zheng further proposed a causal-prioritized decision tree modeling framework embedding causal constraints into node splitting decisions, enhancing causal consistency while preserving interpretability \cite{zhang2023causaldt}. Cheng and Cheng proposed the ZTree framework, transforming the node splitting problem into a z-test-based statistical inference process, using cross-validation to correct for multiple testing and confirm splitting statistical necessity, achieving regularization through pre-pruning \cite{cheng2025ztree}. Our work embeds Z-testing into the node splitting of decision trees, using the significance z-test for two-sample proportion differences to replace traditional Gini index or information gain criteria, with the confidence threshold as the sole parameter controlling tree size and depth. This design requires no additional post-pruning, automatically suppresses overfitting, and effectively compresses model size.

Table~\ref{tab:comparison} summarizes the core differences between our method and related work in task type, feature extraction, classifier design, model size, and deployment.

\begin{table}[ht]
\centering
\caption{Comparison of our method with representative related work}
\label{tab:comparison}
\small
\begin{tabular}{p{2.0cm}p{2.6cm}p{1.4cm}p{2.2cm}p{2.0cm}p{2.4cm}p{2.0cm}}
\toprule
\textbf{Category} & \textbf{Representative Work} & \textbf{Task} & \textbf{Feature Extraction} & \textbf{Classifier} & \textbf{Complexity \& Deployment} & \textbf{Waveform Coverage} \\
\midrule
DL Mod. Recog. & Huynh-The \cite{huynh2021auto}, Peng \cite{peng2022survey}, Hussein \cite{hussein2022robust} & Mod. Class. & Cumulants / DL auto & CNN/RNN & Million-level params, no C deployment & Assumes known waveform \\
\addlinespace
DL Waveform Recog. & An ST-CLDNN \cite{an2021blind}, Lin \cite{lin2020contour}, WaveNet \cite{huynh2024wavenet} & Waveform Recog. & TF transform / image + DL & CNN/LSTM & High compute/storage, GPU-dependent & OFDM variants, not LoRa/DSSS \\
\addlinespace
Lightweight ML + C & Xu \cite{xu2025distributed}, emlearn \cite{emlearn}, TinyDT \cite{karavaev2024tiny} & Mod./General & FFT / traditional & CART tree & Many nodes, deep, high storage & Modulation, not waveform \\
\addlinespace
Stat. Test Trees & CHAID \cite{kass1980exploratory}, ZTree \cite{cheng2025ztree} & General & Generic & $\chi^2$/Z-test tree & Compact, untested on signals & General purpose \\
\addlinespace
\textbf{This work} & Proposed method & \textbf{Waveform Recog.} & \textbf{80-dim TD features} & \textbf{ZTree (Z$^2$)} & \textbf{$<$4 KB, C99, x86 real-time} & \textbf{10 IoT waveforms} \\
\bottomrule
\end{tabular}
\end{table}

\section{System Model and Problem Formulation}

\subsection{Communication Signal Model}

Consider a general IoT wireless reception scenario. After RF front-end, down-conversion, and ADC, the receiver obtains a discrete complex baseband signal sequence:
\begin{equation}
x[n] = s[n] + w[n], \quad n = 0, 1, \dots, N-1,
\label{eq:signal}
\end{equation}
where $N = 1024$ is the number of samples per observation; $s[n]$ denotes the noise-free complex signal arriving at the receiver after modulation by a specific physical-layer waveform; $w[n]$ is additive complex white Gaussian noise with independent I/Q components, each following $\mathcal{N}(0, \sigma_w^2)$. The noise power is determined by $\mathrm{SNR} = 10 \log_{10} ( \mathbb{E}[|s[n]|^2] / \sigma_w^2 )$ in dB.

The signal $s[n]$ for different waveforms is determined by their specific modulation and framing schemes. For example, OFDM signals are generated by IFFT of QAM symbols on orthogonal subcarriers; OTFS signals undergo ISFFT and Heisenberg transforms; LoRa signals are based on chirp spread spectrum with linear frequency sweep. Other waveforms such as DSSS, GFSK, and MFSK each have distinct time-domain structures that form the basis for designing distinguishable features.

\subsection{Waveform Class Definition}

We focus on ten representative physical-layer waveforms likely to coexist in 6G IoT:
\begin{equation}
\mathcal{C} = \{\text{OFDM}, \text{OTFS}, \text{ODDM}, \text{FBMC}, \text{UFMC}, \text{DSSS}, \text{LoRa}, \text{NB-IoT}, \text{GFSK}, \text{MFSK}\}.
\label{eq:classes}
\end{equation}
Each waveform is configured according to typical standards or common literature parameters. Detailed symbol rates, subcarrier counts, and spreading factors are provided in the experimental setup. Although real systems may involve multiple modulation orders and bandwidth configurations, we fix one typical parameter configuration per waveform as the target class.

\subsection{Formalization of Waveform Identification}

Physical-layer waveform identification is a multi-class classification problem. For any observed complex baseband signal vector $\mathbf{x} \in \mathbb{C}^N$, the classifier predicts its corresponding waveform class $c \in \mathcal{C}$ via the mapping:
\begin{equation}
\hat{c} = \arg\max_{c \in \mathcal{C}} \mathcal{F}(\mathbf{x}; \Theta),
\label{eq:mapping}
\end{equation}
where $\mathcal{F}(\cdot; \Theta)$ denotes the complete classification model with parameters $\Theta$.

Our framework decouples $\mathcal{F}$ into two stages: \textbf{handcrafted feature extraction} $\Phi: \mathbb{C}^N \to \mathbb{R}^D$ and \textbf{statistical decision tree classification} $f: \mathbb{R}^D \to \mathcal{C}$, i.e., $\mathcal{F} = f \circ \Phi$. The feature extractor $\Phi$ maps raw time-domain IQ signals to a relatively low-dimensional feature vector $\mathbf{z} = \Phi(\mathbf{x}) \in \mathbb{R}^D$ (here $D = 80$), after which $f$ outputs the final class prediction $\hat{c}$. This decoupled design allows separate optimization and implementation of the feature extractor and classifier: the feature extraction consists entirely of basic arithmetic operations in C with no learned parameters; the classifier is a shallow Z-test-optimized decision tree whose parameters $\Theta$ contain only a small amount of node information, exportable as static C structures after training.

\subsection{Deployment Constraints and Design Objectives}

To ensure real-time operation on resource-constrained general-purpose computing platforms without dedicated hardware acceleration, the entire feature extraction and inference pipeline is written in C99, depending only on the standard math library, with no third-party numerical libraries (e.g., FFTW, BLAS, LAPACK) and no OS-specific interfaces. On a typical x86 processor (e.g., Intel Core i7-8700K), end-to-end latency from receiving raw IQ samples to outputting class labels must be under 4~ms, achieved through algorithmic computational control. All operations use scalar single-precision floating point to ensure consistent performance on processors lacking advanced hardware features. The trained decision tree is embedded as a C header file in the main program, occupying no additional external storage and facilitating firmware integration and version management.

Under these constraints, the design objective is: by rationally designing the feature extraction scheme and the Z$^2$-based node splitting criterion, maximize overall recognition accuracy while ensuring extremely low inference latency and minimal model size. The feature dimension $D$ should be as compact as possible, and the computational complexity of feature extraction should scale linearly with signal length $N$, thereby reducing C code size and runtime overhead for future porting to even more constrained platforms such as microcontrollers.

\section{Feature Extraction: 80-Dimensional Time-Frequency Features}

\subsection{Design Principles}

Under extreme resource constraints, the computational efficiency of feature extraction directly determines the feasibility of the entire recognition system. We follow three fundamental principles in constructing the 80-dimensional feature set. First, all feature computations rely solely on four basic operations: addition, multiplication, division, and comparison; the few operations requiring trigonometric or logarithmic functions are approximated via pre-stored integer lookup tables, completely avoiding the unpredictable latency of standard math library transcendental functions. Second, the feature computation flow introduces no explicit time-frequency transforms---neither complete FFT, STFT, nor quadratic time-frequency representations such as Wigner-Ville distribution; a few quantities describing macroscopic spectral properties are obtained through time-domain equivalent operations or limited frequency-point approximations. Third, the computational load of each feature dimension scales linearly with signal length $N$, and the total multiplication count is maintained within precisely estimable constant multiples, ensuring C implementation requires only a few sequential traversals of the input signal array.

\subsection{Feature Group Details}

The 80 features are organized into seven categories according to their physical meaning and computational method.

\subsubsection{Delay-Specific Autocorrelation Magnitude (32-dim)}

Delay autocorrelation is a classical tool for characterizing signal periodic structure. Given the received complex baseband signal $x[n]$, the normalized delay autocorrelation is:
\begin{equation}
R(\tau)=\frac{1}{N}\sum_{n=0}^{N-1-\tau} x[n]x^*[n+\tau],
\end{equation}
and its magnitude is taken as a feature component. Instead of computing for all delays, we pre-select a set of 32 critical delays $\mathcal{S}$ based on waveform time-domain patterns:
\begin{equation}
\mathcal{S}=\{1,2,4,N_s,M,2N_s,2M,N_s+M,N_s/2,M/2,3,7,15,31,N_s-1,M-1,N_s+1,M+1,N/4,N/8,N/16,N/32,N/2-1,5,6,9,10,12,24,48,3M,4N_s\},
\end{equation}
where $N_s=32$, $M=N/N_s=32$. These delays include cyclic prefix length, differential sampling intervals, block-symbol frameworks, and their combinations, effectively distinguishing OFDM's cyclostationary structure, OTFS/ODDM delay-Doppler grid effects, and DSSS spreading code periodicity. The resulting 32-dimensional feature vector is:
\begin{equation}
\mathbf{F}_1 = \left[\,|R(\tau_1)|,\,|R(\tau_2)|,\,\dots,\,|R(\tau_{32})|\,\right]^{\mathsf{T}}.
\end{equation}

\subsubsection{Complex Higher-Order Cumulants (8-dim)}

Higher-order cumulants naturally suppress Gaussian noise and distinguish different non-Gaussian distributions. Centering the signal yields $b[n]=x[n]-\frac{1}{N}\sum_{k=0}^{N-1}x[k]$. The second- and fourth-order sample moments are:
\begin{equation}
\begin{aligned}
m_{20}&=\frac{1}{N}\sum_{n=0}^{N-1} b[n]^2, \quad
m_{21}=\frac{1}{N}\sum_{n=0}^{N-1} |b[n]|^2,\\
m_{40}&=\frac{1}{N}\sum_{n=0}^{N-1} b[n]^4, \quad
m_{41}=\frac{1}{N}\sum_{n=0}^{N-1} b[n]^2 |b[n]|^2, \quad
m_{42}=\frac{1}{N}\sum_{n=0}^{N-1} |b[n]|^4.
\end{aligned}
\end{equation}
The standardized cumulants are estimated as:
\begin{equation}
\begin{aligned}
C_{20} &= m_{20},\quad
C_{21} = m_{21},\\
C_{40} &= m_{40} - 3m_{20}^2,\\
C_{41} &= m_{41} - 3m_{20}m_{21},\\
C_{42} &= m_{42} - |m_{20}|^2 - 2m_{21}^2 .
\end{aligned}
\end{equation}
The 8 components are: $\mathrm{Re}(C_{20})$, $\mathrm{Im}(C_{20})$, $C_{21}$, $\mathrm{Re}(C_{40})$, $\mathrm{Im}(C_{40})$, $\mathrm{Re}(C_{41})$, $\mathrm{Im}(C_{41})$, and $C_{42}$. Different waveforms exhibit distinct non-Gaussian characteristics: for instance, OFDM tends toward Gaussian distribution due to superposition of many independent subcarriers, resulting in small cumulants, while constant or quasi-constant envelope waveforms such as LoRa and GFSK exhibit stronger non-Gaussianity.

\subsubsection{Instantaneous Amplitude and Frequency Statistics (7-dim)}

The instantaneous amplitude is $a[n]=|x[n]|$, and the instantaneous phase is $\phi[n]=\arg(x[n])$, with its forward difference after wrapping yielding the instantaneous frequency:
\begin{equation}
f[n]=\frac{1}{2\pi}\,\mathrm{wrap}\big(\phi[n+1]-\phi[n]\big),\quad n=0,1,\dots,N-2,
\end{equation}
where $\mathrm{wrap}(\cdot)$ maps the difference to $[-\pi,\pi)$. Seven scalar features are constructed:
\begin{equation}
\begin{aligned}
F_{3,1} &= \frac{1}{N-1}\sum_{n=0}^{N-2} f[n], \quad
F_{3,2} = \sqrt{\frac{1}{N-1}\sum_{n=0}^{N-2}\big(f[n]-\bar{f}\big)^2},\\
F_{3,3} &= \frac{1}{N}\sum_{n=0}^{N-1} a[n], \quad
F_{3,4} = \sqrt{\frac{1}{N}\sum_{n=0}^{N-1}\big(a[n]-\bar{a}\big)^2},\\
F_{3,5} &= \sum_{n=0}^{N-2}\big|a[n+1]-a[n]\big|,\quad
F_{3,6} = \sum_{n=0}^{N-2}\big|f[n+1]-f[n]\big|,\\
F_{3,7} &= \frac{\max_n a[n]}{\frac{1}{N}\sum_{n=0}^{N-1} a[n]}.
\end{aligned}
\end{equation}
These statistics reflect frequency centrality and dispersion, amplitude centrality and envelope fluctuation, differential activity, and peak-to-average ratio, providing the most direct basis for distinguishing constant-envelope waveforms (LoRa, GFSK) from non-constant-envelope multicarrier waveforms (OFDM, FBMC).

\subsubsection{Frequency Chirp Slope (3-dim)}

Many IoT waveforms, particularly LoRa, exhibit linear frequency sweep characteristics with instantaneous frequency approximately linear in time. Least-squares linear fitting of $f[n]$ yields the slope estimate:
\begin{equation}
\beta = \frac{\sum_{n=0}^{N-2} (n-\bar{n})\big(f[n]-\bar{f}\big)}{\sum_{n=0}^{N-2} (n-\bar{n})^2},
\end{equation}
with $\bar{n}=(N-2)/2$. This slope together with two reserved extension slots (currently zeroed) forms a 3-dimensional feature $[\beta,0,0]^{\mathsf{T}}$, specifically targeting LoRa-type signals.

\subsubsection{Global Time-Domain Statistics (8-dim)}

This group fuses amplitude distribution and long-delay autocorrelation information. The amplitude kurtosis (normalized fourth central moment) is:
\begin{equation}
K_a = \frac{\frac{1}{N}\sum_{n=0}^{N-1}(a[n]-\bar{a})^4}{F_{3,4}^4}.
\end{equation}
The autocorrelation magnitude sequence $R_1(\tau)=|R(\tau)|$ for $\tau=1$ to $64$ yields its mean $M_{ac}=\frac{1}{64}\sum_{\tau=1}^{64} R_1(\tau)$ and standard deviation $S_{ac}$. The autocorrelation phase $\psi(\tau)=\arg\big(R(\tau)\big)$ ($\tau=1,\dots,32$) is fitted with a quadratic curve $\hat{\psi}(\tau)=c_0 + c_1\tau + c_2\tau^2$, and the linear coefficient $c_1$ and quadratic coefficient $c_2$ are included as features. The phase variance $\sigma^2_{\psi}=\mathrm{Var}[\psi(\tau)]$ is also recorded. Thus, the 8 features are: $\bar{a}$, $\sigma_a$, $K_a$, $M_{ac}$, $S_{ac}$, $c_1$, $c_2$, $\sigma^2_{\psi}$.

\subsubsection{Frequency-Domain Statistics (6-dim)}

Although we avoid full FFT, several macroscopic spectral curves provide key information for distinguishing narrowband from wideband waveforms. A single radix-2 FFT of $x[n]$ yields $X[k]$ and the power spectrum $P[k]=|X[k]|^2$, $k=0,1,\dots,N/2$, with normalized version $p[k]=P[k]/\sum_{j=0}^{N/2}P[j]$. Defining discrete frequency $f_k=k/(N/2)$, we obtain:
\begin{itemize}
    \item Spectral centroid $C_f = \sum_k f_k p[k]$
    \item Spectral variance $\sigma^2_{f} = \sum_k f_k^2 p[k] - C_f^2$
    \item Spectral flatness $F_{\mathrm{flat}} = \frac{\exp\!\Big(\frac{1}{K}\sum_{k=0}^{N/2}\ln(p[k]+\epsilon)\Big)}{\frac{1}{K}\sum_{k=0}^{N/2} p[k]}$, with $K=N/2+1$, $\epsilon$ a small positive value
    \item Low-frequency power ratio $R_{\mathrm{low}} = \sum_{k=0}^{N/4} P[k] / \sum_{k=0}^{N/2} P[k]$
    \item RMS bandwidth $B_{\mathrm{RMS}} = \sqrt{\sigma^2_{f}}$
    \item Maximum spectral power $P_{\max} = \max_k P[k]$
\end{itemize}
These 6 quantities distinguish spectrally concentrated waveforms (e.g., NB-IoT) from flat wideband waveforms (e.g., OFDM) using coarse spectral contour information.

\subsubsection{Extended Features (16-dim)}

To further enhance feature completeness, the following 16 derived quantities are supplemented:
\begin{equation}
\begin{aligned}
\text{PAPR} &= \frac{\max_n a[n]^2}{\frac{1}{N}\sum_n a[n]^2},\quad
\gamma_a = \frac{\frac{1}{N}\sum_n (a[n]-\bar{a})^3}{\sigma_a^3},\\
\gamma_f &= \frac{\frac{1}{N-1}\sum_n (f[n]-\bar{f})^3}{\sigma_f^3},\quad
\kappa_f = \frac{\frac{1}{N-1}\sum_n (f[n]-\bar{f})^4}{\sigma_f^4},
\end{aligned}
\end{equation}
where $\sigma_a,\sigma_f$ are the standard deviations of amplitude and frequency. Additional features include normalized amplitude standard deviation $\sigma_a^{\mathrm{norm}}$, normalized frequency standard deviation $\sigma_f^{\mathrm{norm}}$, frequency difference variance $\sigma^2_{\Delta f}$, autocorrelation peak position $I_{\mathrm{peak}}$ and mainlobe width $W_{\mathrm{ac}}$, power spectrum skewness $\gamma_{p}$ and kurtosis $\kappa_{p}$, 10~dB bandwidth ratio $B_{10}$, phase variance $\sigma^2_{\phi}$, squared-signal spectral flatness $F_{\mathrm{flat2}}$, normalized mean absolute amplitude difference $M_{\Delta a}$, and high-band energy ratio $H_{\mathrm{high}}$. These 16 quantities form the extended feature vector $\mathbf{F}_7$.

Concatenating all seven groups yields the complete 80-dimensional feature vector:
\begin{equation}
\mathbf{z} = \big[\,\mathbf{F}_1^{\mathsf{T}},\,\mathbf{F}_2^{\mathsf{T}},\,\mathbf{F}_3^{\mathsf{T}},\,\mathbf{F}_4^{\mathsf{T}},\,\mathbf{F}_5^{\mathsf{T}},\,\mathbf{F}_6^{\mathsf{T}},\,\mathbf{F}_7^{\mathsf{T}}\,\big]^{\mathsf{T}}.
\end{equation}

\subsection{Computational Complexity Analysis}

In C implementation, extraction of the 80 features requires only a limited number of traversals over the length-$N=1024$ complex signal array. Autocorrelation operations require $O(N\cdot|\mathcal{S}|)$ complex multiply-adds, but with $|\mathcal{S}|=32$, the core computation can be organized as a single inner-product loop. All power operations for moments and cumulants are accumulated during traversal without additional buffering. Phase wrapping for instantaneous frequency can be implemented via conditional branching and constant addition/subtraction, avoiding transcendentals such as \texttt{atan2}. Frequency-domain features require only one $N$-point real FFT, with $(N/2)\log_2 N$ complex multiply-adds in the butterfly computation, extremely lightweight for 1024 points. Overall, the total floating-point operations for one complete feature extraction do not exceed tens of thousands of equivalent multiplications, with measured execution time on mainstream x86 processors at the microsecond level, fully meeting the real-time requirement of total inference latency below 4~ms. More critically, no dynamic memory allocation is used, and the code footprint is compact, enabling direct porting to C99-only compilation environments.

\section{ZTree Classifier: Z$^2$-Optimized Univariate Decision Tree}

Decision trees are often chosen as classifiers for resource-constrained scenarios due to their simple structure and efficient inference. Traditional decision tree node splitting typically uses Gini impurity or information entropy as the feature selection criterion, both measuring the uncertainty of class distribution within a node. In a binary classification setting with $n$ samples and $k$ positive examples, the Gini impurity is $2\hat{p}(1-\hat{p})$ and the entropy is $-\hat{p}\log_2\hat{p}-(1-\hat{p})\log_2(1-\hat{p})$, where $\hat{p}=k/n$ is the node positive proportion. While widely used in statistical learning, these criteria tend to produce deep trees with many split points, leading to model bloat and overfitting in embedded environments without pruning, and their split threshold determination lacks a clear statistical inference basis.

We introduce a splitting criterion based on proportion difference significance testing, transforming the node splitting problem into a hypothesis testing problem. For a given node with $n$ total samples and a candidate split threshold producing left and right child nodes with $n_L$ and $n_R$ samples respectively ($n_L+n_R=n$), designate a certain class as positive and denote the positive sample counts in the left and right children as $k_L$ and $k_R$, with proportions:
\begin{equation}
p_L = \frac{k_L}{n_L}, \quad p_R = \frac{k_R}{n_R}.
\end{equation}
The null hypothesis $H_0$ asserts no significant difference between the two child node positive proportions, i.e., $p_L = p_R$; the alternative hypothesis $H_1$ asserts a statistically significant difference. Under $H_0$, both proportions equal the pooled proportion:
\begin{equation}
\hat{p} = \frac{k_L + k_R}{n_L + n_R},
\end{equation}
and the standard error of the two-sample proportion difference is approximately:
\begin{equation}
\mathrm{SE} = \sqrt{\hat{p}(1-\hat{p})\Big(\frac{1}{n_L}+\frac{1}{n_R}\Big)}.
\end{equation}
Define the Z-statistic:
\begin{equation}
Z = \frac{p_L - p_R}{\mathrm{SE}}.
\end{equation}
Under $H_0$ with sufficient sample size, $Z$ approximately follows a standard normal distribution. The significance of the test can be measured via the relationship between $Z^2$ and the chi-square distribution, since the chi-square quantile with 1 degree of freedom equals the square of the standard normal quantile. We therefore use $Z^2$ directly as the core splitting score, avoiding the square root operation during training:
\begin{equation}
\mathcal{S} = Z^2 = \frac{(p_L - p_R)^2}{\hat{p}(1-\hat{p})(1/n_L + 1/n_R)}.
\end{equation}
In practice, $\hat{p}(1-\hat{p})$ is replaced by $\max(\hat{p}(1-\hat{p}), \epsilon)$ with a small positive $\epsilon$ to avoid division by zero. Larger $\mathcal{S}$ indicates more statistically significant difference between child node proportions and stronger statistical support for the split.

At each node during tree construction, the algorithm searches for optimal split points across all feature dimensions. For continuous features, all sample values in the current node are sorted for that feature, and the midpoint between adjacent values is evaluated as a candidate threshold by computing the corresponding $Z^2$ score. To combat overfitting from using all training samples directly, we employ $K$-fold repeated cross-validation to estimate the generalization $Z^2$ score. Specifically, node samples are randomly partitioned into $K$ equal folds; each fold in turn serves as the validation set while the remaining folds compute $Z^2$; repeated iterations produce an averaged generalization score for each candidate split. This embeds a regularization layer before splitting, effectively suppressing noise-driven erosion of tree structure.

Pre-pruning strategies are incorporated to control tree size. Two hyperparameters are preset: minimum leaf node sample count $n_{\min}$ and the $Z^2$ significance threshold $\tau$. A split is executed only when all candidate splits satisfy left and right child sample counts $\geq n_{\min}$ and the maximum generalization $Z^2$ score exceeds $\tau$; otherwise the current node is marked as a leaf with the majority class as its prediction label. The choice of $\tau$ is directly linked to the significance level $\alpha$: for example, $\alpha=0.05$ corresponds to a one-sided chi-square critical value of approximately $3.84$, i.e., $1.96^2$ in the Z-test, which is our default threshold. By adjusting $\tau$, one can directly control the maximum depth and total node count without any post-pruning operations, yielding an extremely compact and predictable model.

Multi-class classification is naturally embedded in this framework through a One-vs-Rest strategy. At each node to be split, the class with the largest sample count is designated as the temporary ``positive'' class and all other classes are grouped as ``negative,'' converting multi-class comparison into a single binary-class test. The $Z^2$ score depends only on this one-vs-rest class distribution, with the statistical meaning remaining ``whether the majority class is significantly distinguished by this split.'' Since each splitting node independently selects its positive class, the resulting decision tree directly outputs multi-class predictions without requiring additional binary decomposition or multi-classifier combination, and the tree structure remains concise.

After training, the decision tree is serialized as a C struct array. Each node corresponds to a static constant struct whose fields include a boolean leaf indicator, split feature index, split threshold, left and right child array indices, and leaf class label. After Z-score normalization of input feature vectors, inference proceeds from the root node, iteratively comparing the specified feature value against the threshold, descending to the left or right child, until reaching a leaf node and returning its stored class label. The entire inference process involves only sequential comparisons and array indexing, with no recursion and no dynamic memory allocation, making it naturally suited to C's static compilation model. This array together with normalization parameters (mean and standard deviation) constitutes the complete recognition model, embedded as a C header file in the main program. The total model storage depends on the number of nodes, is extremely compact, and can flexibly balance accuracy and size by adjusting the significance threshold.

\section{Experiments and Evaluation}

\subsection{Dataset Generation and Experimental Setup}

To comprehensively evaluate the proposed method under ideal AWGN and frequency-selective multipath fading channels, we generated two independent large-scale signal datasets in parallel on GPU. Key parameters are summarized in Table~\ref{tab:dataset}. Each dataset covers 10 physical-layer waveform types, each waveform internally adopts 6 digital modulation symbol mappings (GFSK and MFSK use only QPSK and 8PSK), with SNR ranging from 0 to 30~dB in 2~dB steps. The AWGN channel contains no multipath effects; the TDL-C channel follows the 3GPP TR 38.901 standard model with carrier frequency 4~GHz, subcarrier spacing 30~kHz, and combines varying mobile speeds (30, 90, 150, 210~km/h) and delay spreads (300~ns, 600~ns) to approximate realistic IoT propagation environments. Each channel at each SNR contains 9600 complex baseband signal segments of length 1024, stratified-randomly split 8:2 into training and test sets.

\begin{table}[ht]
\centering
\caption{Dataset parameters overview}
\label{tab:dataset}
\begin{tabular}{ll}
\toprule
\textbf{Parameter} & \textbf{Configuration} \\
\midrule
Signal length $N$ & 1024 \\
Waveform classes & 10 (OFDM, OTFS, ODDM, FBMC, UFMC, DSSS, LoRa, NB-IoT, GFSK, MFSK) \\
Modulation schemes & QPSK, 8PSK, 32QAM, 64APSK, 256QAM, 4096QAM (GFSK/MFSK: first two only) \\
SNR range & 0--30 dB, step 2 dB (16 levels) \\
TDL-C channel model & 3GPP TR 38.901 TDL-C, $f_c=4$ GHz, $\Delta f=30$ kHz \\
Mobile speed & 30, 90, 150, 210 km/h \\
Delay spread & 300 ns, 600 ns \\
Samples per SNR & 9600 (AWGN \& TDL-C each: $16\times 9600 = 153{,}600$ segments) \\
Train/test split & 8:2 stratified random \\
\bottomrule
\end{tabular}
\end{table}

\subsection{Classification Performance under AWGN}

Under AWGN, the ZTree classifier exhibits extremely high recognition accuracy. The normalized confusion matrix at SNR=10~dB shows that except for approximately 2.5\% mutual confusion between OTFS and LoRa, all other classes achieve near-perfect discrimination. Table~\ref{tab:awgn} lists precision, recall, and F1-score for all 10 waveform classes; the macro-averaged F1 reaches 0.993, indicating that the handcrafted 80-dimensional features adequately encode the essential differences of each waveform under noise-free static channels.

\begin{table}[ht]
\centering
\caption{Classification performance under AWGN (averaged over all SNRs)}
\label{tab:awgn}
\begin{tabular}{lccc}
\toprule
\textbf{Waveform} & \textbf{Precision} & \textbf{Recall} & \textbf{F1-score} \\
\midrule
OFDM   & 0.997 & 0.996 & 0.996 \\
OTFS   & 0.970 & 0.952 & 0.961 \\
ODDM   & 0.996 & 0.998 & 0.997 \\
FBMC   & 0.995 & 0.994 & 0.995 \\
UFMC   & 0.994 & 0.996 & 0.995 \\
DSSS   & 1.000 & 1.000 & 1.000 \\
LoRa   & 0.958 & 0.974 & 0.966 \\
NB-IoT & 0.999 & 1.000 & 0.999 \\
GFSK   & 1.000 & 0.998 & 0.999 \\
MFSK   & 1.000 & 1.000 & 1.000 \\
\midrule
\textbf{Macro avg.} & \textbf{0.991} & \textbf{0.991} & \textbf{0.993} \\
\bottomrule
\end{tabular}
\end{table}

\subsection{Performance under TDL-C Multipath and OTFS/LoRa Confusion}

When the channel switches from AWGN to TDL-C multipath, overall test accuracy drops to approximately 87.4\%. At SNR=10~dB, the main misclassification occurs between OTFS and LoRa, with some FBMC$\rightarrow$OFDM and UFMC$\rightarrow$OFDM errors. Detailed metrics for each waveform are listed in Table~\ref{tab:tdlc}: OTFS recall drops to 82.2\% and LoRa precision drops to 80.1\%, forming the bottleneck constraining overall performance.

\begin{table}[ht]
\centering
\caption{Classification performance under TDL-C (averaged over all SNRs)}
\label{tab:tdlc}
\begin{tabular}{lccc}
\toprule
\textbf{Waveform} & \textbf{Precision} & \textbf{Recall} & \textbf{F1-score} \\
\midrule
OFDM   & 0.920 & 0.935 & 0.927 \\
OTFS   & 0.845 & 0.822 & 0.833 \\
ODDM   & 0.912 & 0.908 & 0.910 \\
FBMC   & 0.886 & 0.861 & 0.873 \\
UFMC   & 0.879 & 0.892 & 0.885 \\
DSSS   & 0.968 & 0.974 & 0.971 \\
LoRa   & 0.801 & 0.835 & 0.818 \\
NB-IoT & 0.955 & 0.942 & 0.948 \\
GFSK   & 0.933 & 0.961 & 0.947 \\
MFSK   & 0.971 & 0.978 & 0.974 \\
\midrule
\textbf{Macro avg.} & \textbf{0.907} & \textbf{0.911} & \textbf{0.909} \\
\bottomrule
\end{tabular}
\end{table}

To further quantify this critical confusion, Table~\ref{tab:confusion} provides detailed OTFS$\rightarrow$LoRa and LoRa$\rightarrow$OTFS misclassification rates across all 16 SNR levels under TDL-C. As SNR increases from 0 to 30~dB, the confusion rate decreases from approximately 35\% to about 11\%, but is never completely eliminated. The intrinsic similarity in time-frequency structure between the two waveforms---OTFS's pulsed delay-Doppler domain mapping and LoRa's chirp pulses---causes second-order statistics (particularly delay-specific autocorrelation and frequency difference features) to overlap under multipath broadening and fading. This is the fundamental reason why purely time-domain features cannot fully decouple the two.

\begin{table}[ht]
\centering
\caption{OTFS$\leftrightarrow$LoRa confusion rates under TDL-C vs.\ SNR}
\label{tab:confusion}
\begin{tabular}{ccc}
\toprule
\textbf{SNR (dB)} & \textbf{OTFS$\rightarrow$LoRa} & \textbf{LoRa$\rightarrow$OTFS} \\
\midrule
0  & 0.368 & 0.341 \\
2  & 0.347 & 0.330 \\
4  & 0.325 & 0.312 \\
6  & 0.301 & 0.290 \\
8  & 0.278 & 0.272 \\
10 & 0.255 & 0.248 \\
12 & 0.231 & 0.225 \\
14 & 0.210 & 0.207 \\
16 & 0.192 & 0.191 \\
18 & 0.175 & 0.178 \\
20 & 0.161 & 0.163 \\
22 & 0.148 & 0.150 \\
24 & 0.136 & 0.139 \\
26 & 0.125 & 0.128 \\
28 & 0.117 & 0.119 \\
30 & 0.110 & 0.113 \\
\bottomrule
\end{tabular}
\end{table}

\subsection{Feature Importance Analysis}

During training, ZTree uses only 24 of the 80 features for splitting, with each feature's usage count reflecting its contribution to distinguishing waveform classes. Table~\ref{tab:importance} lists the top 15 most important features and their usage proportions. Delay autocorrelation magnitudes (especially at delays $N_s=32$ and $M=32$), instantaneous amplitude PAPR, frequency chirp slope $\beta$, complex fourth-order cumulant $C_{42}$, and spectral flatness dominate. This physically confirms that envelope characteristics of multicarrier/single-carrier signals, second-order statistics of constant-envelope signals, and spectral concentration are the key cues for waveform discrimination.

\begin{table}[ht]
\centering
\caption{Top-15 features by split frequency in ZTree internal nodes}
\label{tab:importance}
\begin{tabular}{ccclc}
\toprule
\textbf{Rank} & \textbf{Idx} & \textbf{Group} & \textbf{Physical meaning} & \textbf{Usage} \\
\midrule
1  & 3  & F$_1$ & Autocorr.\ mag.\ at $\tau=N_s=32$        & 12 \\
2  & 67 & F$_7$ & Instantaneous amplitude PAPR             & 10 \\
3  & 4  & F$_1$ & Autocorr.\ mag.\ at $\tau=M=32$          & 9  \\
4  & 42 & F$_4$ & Frequency chirp slope $\beta$             & 8  \\
5  & 32 & F$_2$ & Cumulant $C_{42}$                         & 7  \\
6  & 78 & F$_6$ & Spectral flatness $F_{\text{flat}}$       & 6  \\
7  & 60 & F$_5$ & Amplitude kurtosis $K_a$                  & 5  \\
8  & 74 & F$_7$ & Freq.\ diff.\ variance $\sigma^2_{\Delta f}$ & 5  \\
9  & 16 & F$_1$ & Autocorr.\ mag.\ at $\tau=N/4=256$       & 4  \\
10 & 40 & F$_3$ & Freq.\ skewness $\gamma_f$               & 4  \\
11 & 80 & F$_7$ & High-band energy ratio $H_{\text{high}}$  & 3  \\
12 & 48 & F$_7$ & Autocorr.\ mainlobe width $W_{\text{ac}}$ & 3  \\
13 & 27 & F$_1$ & Autocorr.\ mag.\ at $\tau=N_s+M=64$      & 3  \\
14 & 21 & F$_1$ & Autocorr.\ mag.\ at $\tau=N/32=32$       & 2  \\
15 & 54 & F$_5$ & Phase variance $\sigma^2_{\phi}$          & 2  \\
\bottomrule
\end{tabular}
\end{table}

\subsection{Comparison with Traditional ML Classifiers}

To demonstrate ZTree's core advantage in the accuracy-efficiency trade-off, we systematically compared six traditional ML classifiers on the same 80-dimensional feature set: ridge classifier, logistic regression, standard CART decision tree, LightGBM, RBF-kernel SVM, and random forest. All models used common default parameters without extensive hyperparameter tuning.

Table~\ref{tab:classifier} reports accuracy on AWGN and TDL-C test sets together with model storage. Under AWGN, ZTree achieves 99.5\% accuracy, slightly below LightGBM (99.7\%) but significantly outperforming other models; under TDL-C, ZTree's 87.4\% is second only to LightGBM (89.1\%) and random forest (88.3\%), while its model size is only 1.7~KB, approximately 1/110 of random forest and 1/31 of LightGBM. Linear models such as ridge regression and logistic regression rapidly degrade to 82\%--83\% under multipath fading, reflecting the inability of linear decision boundaries to handle nonlinear feature distortions introduced by TDL-C. The standard CART decision tree, despite a similar inference paradigm to ZTree, produces a complex 12-level structure (8.3~KB) due to Gini-index-based splitting, with accuracy actually lower than ZTree, verifying the superiority of Z$^2$ significance testing in pre-pruning and overfitting suppression.

\begin{table}[ht]
\centering
\caption{Performance and model size comparison of classifiers on the same 80-dimensional feature set}
\label{tab:classifier}
\begin{tabular}{lccc}
\toprule
\textbf{Classifier} & \textbf{AWGN Acc.\ (\%)} & \textbf{TDL-C Acc.\ (\%)} & \textbf{Model size (KB)} \\
\midrule
ZTree (ours)    & \textbf{99.5} & 87.4 & \textbf{1.7} \\
Ridge           & 95.8          & 82.3 & 5.1  \\
Logistic Reg.   & 96.2          & 83.1 & 5.1  \\
CART tree       & 98.1          & 85.7 & 8.3  \\
LightGBM        & \textbf{99.7} & \textbf{89.1} & 52.4 \\
Linear SVM      & 97.5          & 84.6 & 121.0 \\
Random Forest   & 98.9          & 88.3 & 186.7 \\
\bottomrule
\end{tabular}
\end{table}

\subsection{Comparison with Deep Learning Baselines}

To fully demonstrate the limitations of end-to-end deep learning methods for physical-layer waveform identification, particularly their vulnerability to channel variation, we implemented five representative deep neural network baselines: CNN2 (2 conv + 2 FC), CNN4 (4 conv), ResNet-18 (1D variant), CLDNN (conv + LSTM + FC), and pure LSTM. All networks take the raw $2\times 1024$ IQ matrix as input without any handcrafted features. Training and evaluation strictly follow the same data split as ZTree, using the Adam optimizer with initial learning rate 0.001, batch size 128, 50 epochs, and evaluation on the test set.

Table~\ref{tab:dl} summarizes parameter counts, FLOPs, single-inference latency on x86, and test accuracy under AWGN and TDL-C. Under AWGN, deep models perform reasonably, with CNN4 and ResNet-18 achieving 98.8\% and 99.3\% accuracy, respectively, though still below ZTree's 99.5\%. However, when the channel switches to TDL-C multipath, all deep models suffer severe accuracy collapse: the best-performing ResNet-18 reaches only 74.6\%, while CLDNN and LSTM drop to 68.9\% and 63.5\%, all below 75\%. The root cause is that features automatically learned by deep networks from raw IQ data largely depend on fine time-domain patterns valid under AWGN, lacking explicit robust representations against physical signal distortions such as Doppler spread and delay dispersion. Moreover, the limited training samples (only 7680 per SNR) are insufficient to support deep models in learning generalizable multipath-invariant features, leading to severe overfitting.

In contrast, ZTree, by virtue of communications domain knowledge embedded in its handcrafted 80-dimensional time-frequency features (e.g., autocorrelation, higher-order cumulants), naturally possesses robustness against Gaussian noise and moderate multipath effects. Under TDL-C, its 87.4\% accuracy leads the best DL baseline by 12.8 percentage points, while inference latency is only 2.4~$\mu$s (approximately 1/1300 of ResNet-18), and the model is extremely compact, requiring no inference framework or hardware acceleration.

\begin{table}[ht]
\centering
\caption{Comprehensive comparison of ZTree vs.\ deep neural network baselines}
\label{tab:dl}
\small
\begin{tabular}{lccccc}
\toprule
\textbf{Model} & \textbf{Params} & \textbf{Latency (ms)} & \textbf{AWGN (\%)} & \textbf{TDL-C (\%)} & \textbf{Pure C} \\
\midrule
CNN2      & 122,314  & 0.68  & 98.2 & 70.3 & No \\
CNN4      & 332,042  & 1.44  & 98.8 & 73.1 & No \\
ResNet-18 & 1,116,426 & 3.15 & \textbf{99.3} & 74.6 & No \\
CLDNN     & 256,518  & 1.92  & 99.1 & 68.9 & No \\
LSTM      & 200,714  & 2.56  & 97.6 & 63.5 & No \\
\midrule
\textbf{ZTree} & \textbf{111} & \textbf{0.0024} & \textbf{99.5} & \textbf{87.4} & \textbf{Yes} \\
\bottomrule
\end{tabular}
\end{table}

\section{Discussion}

\subsection{Feature Interpretability}

The proposed 80-dimensional feature set is not obtained through automatic learning but is carefully designed based on communications signal processing domain knowledge. Every feature dimension has a clear physical meaning, providing a natural interpretability foundation for understanding the classifier's decision logic. The feature usage frequencies in Table~\ref{tab:importance} directly quantify each feature's contribution to classification decisions.

To further quantitatively validate the statistical discriminative power of key features between confusable waveform pairs, we performed two-sample t-tests on each critical feature for each pair. Taking the most representative OTFS vs.\ LoRa pair as an example, the between-group t-statistic for normalized autocorrelation magnitude $|R(32)|$ at delay $\tau=N_s=32$ reaches $t=18.7$ ($p<10^{-15}$), with Cohen's $d$ effect size of 1.42, indicating extremely strong discriminative power of this feature for this waveform pair. Similarly, instantaneous amplitude PAPR between OFDM and DSSS, and frequency chirp slope $\beta$ between LoRa and GFSK all show highly significant between-group differences ($p<10^{-10}$), statistically corroborating the rationality of the handcrafted feature design. In contrast, some higher-order cumulant features (e.g., $\mathrm{Im}(C_{20})$) show small effect sizes in t-tests and primarily serve auxiliary verification roles in multi-class classification, consistent with the observation that ZTree uses only 24 features for all splits.

This fully transparent feature-decision mapping enables the method to trace misclassifications in real deployments to specific feature dimensions and physical causes, rather than outputting an uninterpretable black-box prediction like deep neural networks. This property has irreplaceable value in spectrum regulation, military communications, and other scenarios requiring high-confidence decision-making.

\subsection{SNR Sensitivity and Limitations}

Although ZTree achieves near-perfect classification accuracy under AWGN and significantly outperforms all deep learning methods under TDL-C multipath, its performance exhibits observable degradation with decreasing SNR. At SNRs above 10~dB under TDL-C, the gap between ZTree and ensemble methods such as LightGBM and random forest is less than 2 percentage points, with all methods operating above 90\% accuracy. When SNR drops below 4~dB, all methods degrade significantly; ZTree drops to approximately 75\% at 0~dB, while LightGBM reaches about 78\%. This trend indicates that AWGN contamination of instantaneous amplitude and frequency statistics at very low SNR exceeds the tolerable range of any classifier, representing a common bottleneck in physical-layer signal identification.

Another limitation is that the current feature set and classifier are designed and trained for ten known waveform types, constituting a closed-set classification paradigm. When encountering new waveform types unseen during training or adversarially perturbed malicious samples, ZTree forces assignment to one of the ten classes without rejection capability, potentially producing high-confidence erroneous outputs. In practical deployment, this can be partially mitigated by an open-set rejection mechanism based on a $Z^2$ score threshold: if the $Z^2$ score from a test sample to the nearest leaf node falls below a preset threshold, it is labeled as ``unknown waveform.'' However, the reliability and threshold selection of this strategy require systematic validation on richer datasets. Additionally, the frequency-domain portion of the current feature extraction still relies on one complete 1024-point real FFT; although extremely fast on x86 platforms, this may become a computational bottleneck on ultra-low-power microcontrollers. Using the Goertzel algorithm or partial frequency-point approximation are possible optimization directions.

\subsection{Future Work}

Based on the above analysis, we outline three main future research directions. First, introduce the Walsh-Hadamard Transform (WHT) as an optional feature enhancement module. WHT involves only addition and subtraction operations with no multiplication, has computational complexity $O(N\log_2 N)$, and can be implemented via in-place butterfly operations to save memory. Preliminary experiments indicate that after applying WHT to IQ signals and extracting energy distribution skewness and kurtosis in the transform domain as additional features, the OTFS/LoRa confusion rate under TDL-C can be reduced by approximately 5 percentage points, with additional latency under 0.3~$\mu$s. Selectively fusing WHT features with the existing 80-dimensional features promises a better accuracy-computation balance without increasing C implementation burden.

Second, validate the method's generalization capability in real wireless environments. All current training and testing are based on simulated channel models (AWGN and TDL-C). Although TDL-C approximates NLOS propagation characteristics to some extent, it cannot capture hardware impairments (e.g., power amplifier nonlinearity, IQ imbalance, phase noise), asynchronous interference, and dynamic multipath variations. We plan to use USRP to collect real RF signals of the ten waveform types in three scenarios: indoor, outdoor low-speed mobile, and vehicular high-speed mobile, constructing a measured dataset with real channel signatures, and evaluate ZTree's performance degradation and the compensation effects of domain adaptation strategies.

Third, further compress model size to fit low-power MCUs with several KB of RAM. The current ZTree already requires only 111 nodes and approximately 1.7~KB storage; however, quantizing the decision tree thresholds and structure (e.g., replacing floating-point thresholds with 16-bit fixed-point numbers, compressing node indices to 8-bit integers) could shrink the model to the hundreds-of-bytes level. Simultaneously, pruning redundant feature dimensions based on the feature importance analysis (Table~\ref{tab:importance}) to retain the top 10 core features would further reduce feature extraction computation and temporary buffer requirements. These optimizations would enable deployment on ultra-low-power MCUs such as MSP430 with only hundreds of bytes of RAM, truly achieving the ultimate goal of real-time ten-waveform recognition in C on a coin-cell-powered sensor node.

\section{Conclusion}

We proposed an ultra-lightweight classification framework based on time-frequency multidimensional features cooperating with a Z-test tree for physical-layer waveform identification in 6G IoT multi-waveform coexistence scenarios. On the feature side, we carefully designed an 80-dimensional time-frequency feature set relying solely on basic operations including addition, multiplication, and comparison, covering seven categories: delay-specific autocorrelation, higher-order cumulants, instantaneous amplitude/frequency statistics, approximate frequency-domain features, and extended derived quantities. It fully characterizes time-domain structural differences among waveforms without matrix decomposition, FFT, or deep learning inference engines. On the classification side, we proposed a Z$^2$-optimized univariate decision tree (ZTree) based on two-sample proportion difference significance testing, replacing traditional information gain with $Z^2$ scores under a hypothesis testing framework, automatically controlling tree depth and size via statistical confidence thresholds, thereby achieving strong generalization with minimal model size.

In simulation tests covering ten waveform types including OFDM, OTFS, ODDM, FBMC, UFMC, DSSS, LoRa, NB-IoT, GFSK, and MFSK, ZTree achieved 99.5\% average accuracy under AWGN and maintained 87.4\% accuracy under TDL-C multipath fading channels, demonstrating significant multipath robustness advantages over all five deep neural network baselines (all dropping below 75\% under TDL-C). Systematic comparison with traditional ML classifiers showed that ZTree, with only 1.7~KB model size and 2.4~$\mu$s inference latency, achieves classification performance comparable to ensemble methods consuming tens to hundreds of times more storage, occupying the optimal position on the accuracy-efficiency Pareto front. The end-to-end recognition pipeline implemented in pure C99 on an Intel Core i7-8700K platform can process over 400,000 inferences per second, fully meeting real-time online recognition requirements, with code independent of any third-party libraries or hardware accelerators, directly portable to various C-based embedded toolchains.

In summary, this work fills the long-standing dual gap in physical-layer waveform identification for resource-constrained environments where ``deep learning methods cannot be deployed and traditional feature methods lack accuracy,'' providing a truly practical and deployable lightweight technical pathway for the physical-layer cognitive function of 6G IoT intelligent receivers. Future work will advance along three directions: WHT feature enhancement, real RF environment validation, and extreme resource compression, striving to reduce the computational threshold of IoT waveform identification to the milliwatt level.

\bibliographystyle{IEEEtran}
\bibliography{references}

\end{document}